\documentclass[a4paper]{VisionStyle}
\usepackage{epsfig}

\begin{document}

\def\1034{RE~J1034$+$396} 
\def\simlt{\lower.5ex\hbox{$\; \buildrel < \over \sim \;$}}
\def\simgt{\lower.5ex\hbox{$\; \buildrel > \over \sim \;$}}

\title{Do NLS1s and ultrasoft AGN have irradiated warped accretion disks?}

\author
    {E. M. \,Puchnarewicz 
\and R. \,Soria
 } 


\institute{
Mullard Space Science Laboratory, University College London, Holmbury
St. Mary, Dorking, Surrey RH5 6NT, UK
}

\maketitle 

\begin{abstract}
When Beppo-SAX measured the 0.1 to 12 keV spectrum of RE~J1034+396,
observations in the optical, UV and EUV were also taken within a few
weeks. This multiwavelength spectrum placed very strong constraints on
its unusually hot big blue bump component, which has been attributed
to a high (almost Eddington) rate of accretion onto a disk surrounding
a low mass black hole. However, while a simple, geometrically-thin
accretion disk provides a good fit to the UV to X-ray continuum, it
leaves a residual flat and featureless component in the optical. We
propose that the disk is actually flared or warped and that the
central ionizing EUV/X-ray continuum is irradiating the outer parts of
the disk and boosting the optical/UV continuum flux. The relatively
narrow permitted lines may be due to a resulting disk wind. This
physical interpretation may explain the link between ultrasoft X-ray
excesses and broad line velocity in the AGN class as a
whole. It may also have implications for the possible relationship 
between ultrasoft AGN and galactic black holes.

\end{abstract}

\section{Introduction}
\label{epuchnarewicz-C2-37_sec:intro}
  
The optical to X-ray continuum of RE~J1034+396 is highly unusual for a
Seyfert 1 galaxy. The optical/UV continua of most AGN rise towards the
blue with a slope $\alpha\sim$~0.4 (where $\alpha$ is the spectral
index, defined such that F$_\nu\propto\nu^{-\alpha}$), and the
soft X-ray spectrum falls towards high energies with spectral index
$\alpha\sim$~2 (e.g. \cite{epuchnarewicz-C2-37:laor97}). The spectrum
seems to peak in the unobservable EUV and this continuous, optical to
soft X-ray feature known as the `big blue bump' (BBB), is believed to
represent the emission from a geometrically thin, optically-thick
accretion disc (AD).

In REJ~1034+396 however, the optical/UV continuum is flat
($\alpha\sim$1) with no sign of the BBB down to Ly$\alpha$. At
$\sim$0.1keV, the soft X-ray spectrum is very strong above the
extrapolated level of the optical/UV continuum, peaking at
$\sim$0.3keV
(\cite{epuchnarewicz-C2-37:puch98}).

\begin{figure}[ht]
  \begin{center}
    \epsfig{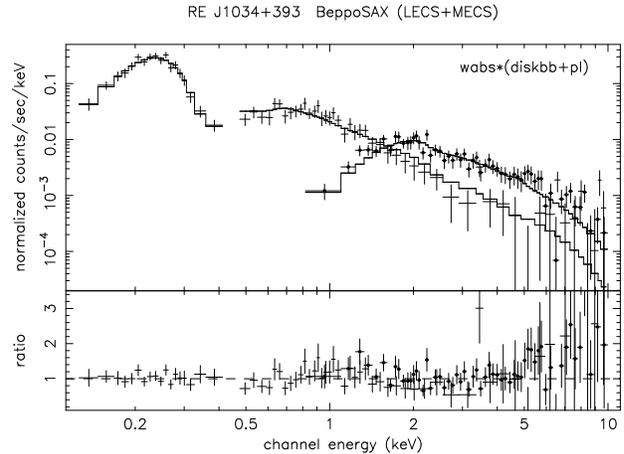}
  \end{center}
\caption{Disk blackbody plus power law fit 
to the 0.1--10 keV {\it Beppo-SAX} spectrum.
}
\label{epuchnarewicz-C2-37_fig:sax}
\end{figure}

The Beppo-SAX spectrum (0.1 to 10~keV) is shown in
Figure~\ref{epuchnarewicz-C2-37_fig:sax}. It was best-fit by
\cite{epuchnarewicz-C2-37:soria} using a Comptonized disk blackbody
with a temperature at the inner edge of the disk, T$_{\rm in}$=120eV
and a coronal temperature T$_{\rm C}$=15keV.

\begin{figure}[ht]
  \begin{center}
    \epsfig{file=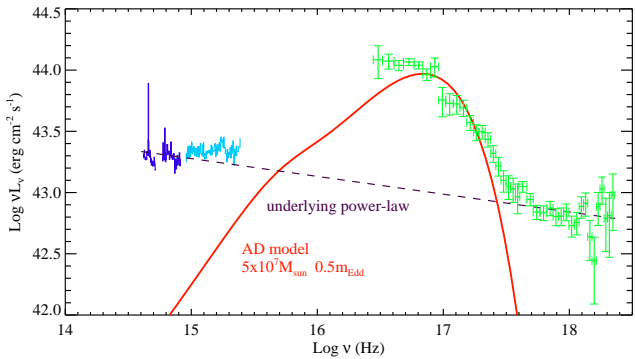, width=9cm}
  \end{center}
\caption{The multiwavelength spectrum of REJ~1034+396 (optical in dark
blue; HST-UV in light blue; Beppo-SAX in green) compared with
the best-fitting AD model (solid red) and power-law component (black
dashed).
}
\label{epuchnarewicz-C2-37_fig:obs}
\end{figure}

The multiwavelength (8000~\AA\ to 10~keV) spectrum, comprising
quasi-simultaneous optical, UV and X-ray data, was measured in January
1997 and fitted with a combination of blackbody and power-law models
(\cite{epuchnarewicz-C2-37:puch01}). A simple geometrically-thin AD was
not sufficient to reproduce this spectrum so an underlying optical to
X-ray power-law component with $\alpha\sim$~0.6 was added
in. \cite{epuchnarewicz-C2-37:alice} investigated the possibility of a
BL~Lac-type of component for this power-law but no compelling evidence
for the physical origin of the 'power-law' was found. The
multiwavelength spectrum used in the fits is shown in
Figure~\ref{epuchnarewicz-C2-37_fig:obs}. The AD and power-law
components are plotted separately so that the shortfall in the
optical/UV can be clearly seen. The fits inferred a high accretion
rate (L~0.3-0.7L$_{\rm Edd}$), a small black hole mass, M$\sim
10^6$M$_\odot$ and a viewing angle of 60-70$^\circ$. Thus they
concluded that RE~J1034+396 had a low-mass black hole accreting close
to the Eddington limit. 

\section{A warped, or flared, accretion disk}
\label{epuchnarewicz-C2-37_sec:warped}

\begin{figure*}[ht]
  \begin{center}
    \epsfig{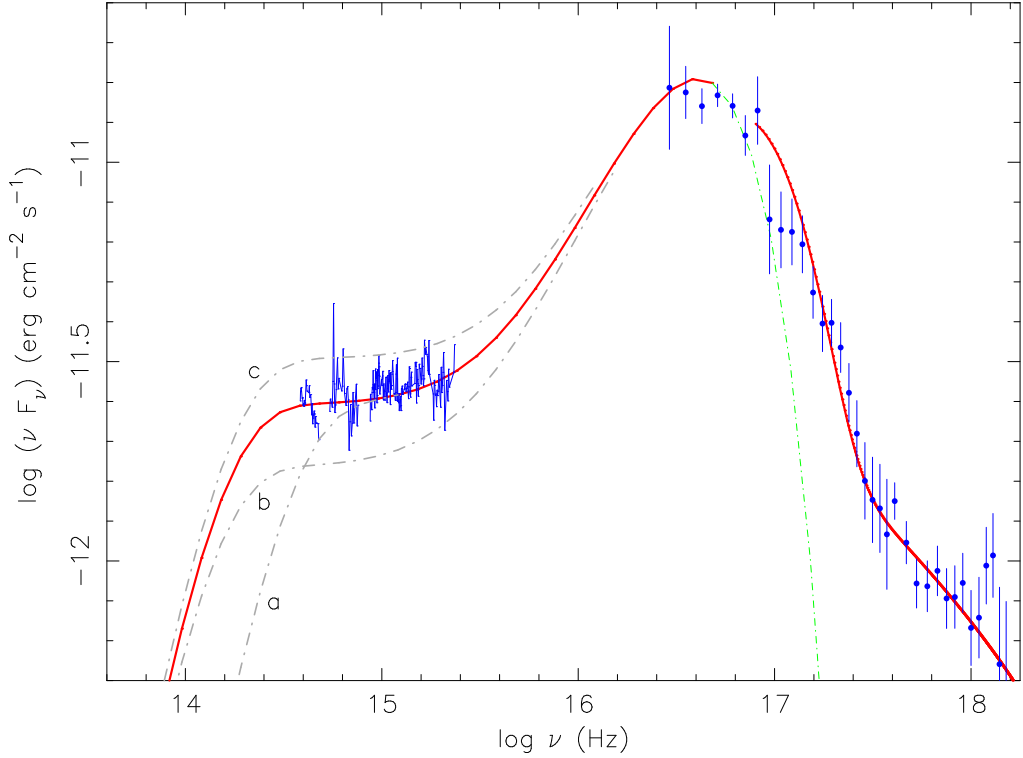}
  \end{center}
\caption{
Broadband spectral fit to the WHT, HST, and {\it Beppo-SAX} 
observed fluxes ($E \leq 7$ keV). 
(The line-of-sight Galactic absorption has been removed.)
The thick solid line for $\log \nu < 16.7$ represents 
the low-energy part of an irradiated disk blackbody spectrum, continued as 
a dash-dotted line for $\log \nu > 16.7$. The thick solid line for 
$\log \nu > 16.9$ represents the X-ray spectral fit, physically 
interpreted as a disk blackbody modified by Comptonisation. 
The temperature at the inner edge of the disk 
$kT_{\rm in} = 0.1$ keV.  
The meaning of the curves labelled 'a', 'b' and 'c' 
is discussed in the text.
}
\label{epuchnarewicz-C2-37_fig:models}
\end{figure*}

The effective temperature of an AD photosphere at each radius 
is determined by the thermal energy generated by local viscous dissipation 
near the disk mid-plane (viscous heating), 
as well as by the energy intercepted 
from the central X-ray source, thermalised and re-emitted 
(irradiative heating). 

In a geometrically-thin, optically-thick AD, at a given
radius R, the temperature is proportional to
R$^{-3/4}$. However, if the disk is irradiated, then the temperature,
T$_{\rm irr}$, is proportional to R$^{-1/2}$ and irradiative heating
will dominate over viscous heating at large radii. If the disk is
flared or warped then emission due to irradiation will be relatively
strong in the optical/UV, where it intercepts more of the ionizing
continuum. The spectrum will then flatten at these wavelengths. This
provides a straightforward solution to the origin of the flat
optical/UV continuum.

The continuum spectrum at energies $\simlt 0.3$ keV can be well fitted
by a diluted disk blackbody spectrum [thick (red) solid line in
Figure~\ref{epuchnarewicz-C2-37_fig:models}] with a
spectral hardening factor, $f = 1.5$ ($f$=1 for an AD which is {\sl not}
irradiated), and a temperature at the inner
edge of the AD of $\sim$100~eV (see Table 2 in
\cite{epuchnarewicz-C2-37:soria} for full details of the best-fit
parameters). The best-fit model is shown in
Figure~\ref{epuchnarewicz-C2-37_fig:models}. Irradiation is dominating
at a radius, R$_{\rm irr}\simgt$~80~R$_{\rm inn}$. For models b and c
in Figure~\ref{epuchnarewicz-C2-37_fig:models}, irradiation dominates
at R$_{\rm irr}\simgt$~110~R$_{\rm inn}$ and R$_{\rm
irr}\simgt$~50~R$_{\rm inn}$ respectively.

The outer disk radius is also constrained by our fit: we require
$R_{\rm out} \simgt 5 \times 10^{16}$ cm $\approx 3 \times 10^5$
$GM/c^2$. If we keep all other parameters in the best-fit model fixed,
but assume a disk truncated at $R_{\rm out} = 10^{16}$ cm, we cannot
reproduce the flat optical spectrum (curve marked 'a' in
Figure~\ref{epuchnarewicz-C2-37_fig:models}).

The
black hole mass, M$\sim 10^6$~M$_\odot$. At this mass, the value of
the innermost stable orbit from this fit, R$_{\rm inn}$, is consistent
with a Kerr black hole.  The fits allow a range of viewing angles so
that the disk may be viewed close to the disk axis (up to $\sim
20^\circ$ away). This removes the problem of having to view the AD
through a co-planar molecular torus and is consistent with suggestions
that ultrasoft AGN are seen relatively face-on
(\cite{epuchnarewicz-C2-37:usoft}).

\section{Emission lines from the disk?}
\label{epuchnarewicz-C2-37:emlines}

A large accretion disk heated by soft X-rays intercepted from the
central source is likely to have a temperature-inversion layer at its
surface. Therefore, we expect strong emission-line formation at large
radii. In particular, in the case of \1034, the largest contribution
to low-ionisation lines such as H$\alpha$ and H$\beta$ would come from
radii $\simgt 10^{16}$ cm, where the Keplerian rotational velocities
are $(GM/R)^{1/2} \simlt 1000$ km s$^{-1}$. If the Balmer lines are
indeed produced near the disk surface, we expect full widths at half
maximum $\sim 2(\sin i) (GM/R)^{1/2} \simlt 2000$ km s$^{-1}$.  This
is in agreement with the observed values of 1500 and 1800 km s$^{-1}$
for H$\beta$ and H$\alpha$ respectively
(\cite{epuchnarewicz-C2-37:puch98}). 

Moreover, if the X-ray luminosity from the central source is $\simgt
0.1$ $L_{\rm Edd}$, we may expect the formation of a
radiatively-driven accretion disk wind.  Accurate modelling of the
emission-line profiles may help us ascertain if higher-ionisation
lines are formed in the wind, and determine the ionisation parameter
of the emission regions. The narrow, single-peaked UV lines seen by
\cite{epuchnarewicz-C2-37:puch98} appear qualitatively consistent with
the disk-wind model of \cite{epuchnarewicz-C2-37:murray}.  The
presence of broad and narrow components in the UV lines may be
explained with a broad component emitted near the irradiated disk
surface at smaller radii, and a narrow component emitted in the
photoionised wind.

\section{Summary}

\begin{figure}[ht]
  \begin{center}
    \epsfig{file=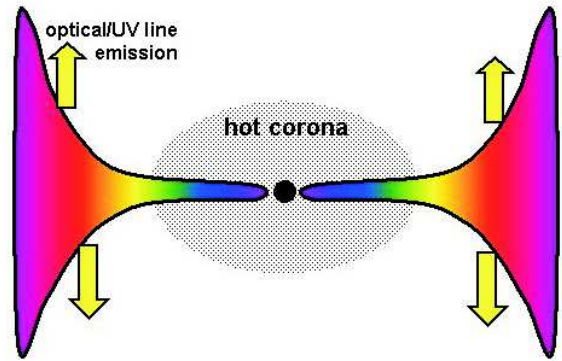, width=8cm}
  \end{center}
\caption{A cross-section through the nucleus of RE~J1034+396. The disk is shown flared here but may also be warped. The cool, outer regions of the disk intercept a greater fraction of the hot, ionizing EUV/X-ray continuum than in a flat disk, enhancing the optical continuum emission. Permitted lines may be emitted from a radiatively-driven disk wind.
}
\label{epuchnarewicz-C2-37_fig:cartoon}
\end{figure}

RE~J1034+396 is an ultrasoft X-ray AGN with relatively narrow
permitted lines. Its accretion disk component is unusually hot
($\sim$100~eV) and shifted out of the UV completely, leaving bare a
flat optical/UV continuum. We find that this component is consistent
with the irradiation of a flared or warped disk. Optical and UV lines
may be produced in a radiatively-driven disk wind and, qualitatively,
their profiles and velocities are consistent with those observed in
RE~J1034+396 and other ultrasoft Seyferts. A sketch of the geometry is
shown in Figure~\ref{epuchnarewicz-C2-37_fig:cartoon}. If RE~J1034+396
is a template for all ultrasoft AGN, it may be that the degree of AD
flaring is a consequence of the extreme ultrasoft X-ray flux and
temperature. The tendency for ultrasoft AGN to have relatively low
emission line velocities (ie they tend to be narrow-line Seyfert 1s:
\cite{epuchnarewicz-C2-37:usoft}) 
may also be due to irradiation of a flared AD, where the emission line
flux is dominated by an AD wind. 

An analogy has already been drawn between ultrasoft AGN and galactic
black holes which share a black hole plus accretion disk
geometry. Such a self-consistent picture in AGN has important
implications for our understanding of black hole physics on mass
scales from stellar to supermassive.

\end{document}